\documentclass[prb,aps,amsfonts,groupedaddress,notitlepage,nofootinbib,longbibliography,twocolumn,10pt]{revtex4-2}
\usepackage{graphicx}
\usepackage[dvipsnames]{xcolor}

\usepackage[colorlinks={true}, 
linkcolor={BrickRed},     % Code: C43119
citecolor={MidnightBlue}, % Code: 006896
urlcolor={PineGreen},     % Code: 058D74
bookmarks={true}, pdftitle={GOE noise}, pdffitwindow={true}, pdfpagetransition =
{Fade}, pdfcenterwindow={true} ]{hyperref} 

\usepackage{amsmath,amssymb,amsfonts}  % Math symbols
\usepackage{graphicx}              % Figures
\usepackage{physics}                % Brackets, derivatives, etc.
\usepackage{bm}                     % Bold math symbols
\usepackage{ragged2e}
\usepackage{booktabs}  % for nicer lines
\usepackage{multirow}  % for merging rows
\usepackage{comment}
\usepackage{xcolor}
\usepackage[normalem]{ulem}
\begin{document}
% \linenumbers

\title{Measuring intrinsic relaxation rates in superconductors using nonlinear response}
\author{Wei-En Tseng and Rahul Nandkishore}
\affiliation{Department of Physics and Center for Theory of Quantum Matter, University of Colorado Boulder, Boulder, Colorado 80309, USA}
\date{\today}

\begin{abstract}
We discuss intrinsic relaxation rates in superconductors, and how they may be measured using non-linear optical (terahertz) response. We consider both $s$ and $d$-wave superconductors, both with and without a phenomenological (energy dependent) damping. Intrinsic relaxation rates of interest include the Higgs mode decay rate, the quasiparticle redistribution rate ($1/T_1$) and the quasiparticle dephasing rate ($1/T_2$), where the latter two rates are zero in the pure BCS model, but non-zero in the presence of damping. Using the Anderson pseudospin formalism, we illustrate how these intrinsic relaxation rates are related to measurable quantities such as the time-dependent gap function and the non-linear current (a.k.a. third harmonic generation). Hence, we show how intrinsic relaxation rates may be experimentally extracted and discuss what one may thereby learn about the underlying damping. We also discuss the effects of polarization control (viz. non-linear response to light polarized in different directions), which offers a useful experimental knob, especially for $d$-wave superconductors, enabling selective excitation of modes in different irreducible representations (and readout of their corresponding relaxation rates).
\end{abstract}

\maketitle
% \tableofcontents

\section{Introduction}

Experimental developments in nonlinear spectroscopy in the optical and terahertz range have opened new pathways to the interrogation of quantum materials. For example, non-linear spectroscopy has been used to probe the Higgs mode in superconductors (see e.g. \cite{Shimano2020} and references contained therein), intrinsic relaxation rates in Silicon \cite{MahmoodSilicon}, and energy relaxation in strange metals \cite{StrangeEnergy}, and have been proposed to offer a powerful probe of fractionalization and spin liquids \cite{WanArmitage, NandkishoreChoiKim, HartNandkishore, McGinleyetal}. In this manuscript, we will examine the question of how such techniques can be used to learn about intrinsic relaxation rates in (clean) superconductors. We will focus on the interrogation of systems that are already superconducting in equilibrium, and will not consider e.g. optically induced superconductivity \cite{Cavalieristuff}. 

Our work builds on an extensive prior literature. For example, it is by now well established that the Higgs (amplitude) mode of a superconductor may be induced via nonlinear response \cite{Shimano2020, Higgs-oscillations-in-time-resolved, Matsunaga, Katsumi, schwarz_classification_2020}. The time dynamics of a superconductor following a {\it quench} generated by an optical pulse, has also been well studied e.g. using the Anderson pseudospin method in \cite{Twisting}, and it is known that coherent `Landau' damping of the resulting oscillations produces a gap function that depends on time according to a universal power law function \cite{Collisioness, gurarieprl, OrthFernandes2}. This analysis was further enriched by adding a phenomenological incoherent damping in \cite{Impact_of_damping} and examining the subsequent time dynamics of the gap function. This paper represents our point of departure. It is also important to note that in THz optical experiments, the measurable quantity is the non-linear current, rather than the Higgs oscillation itself \cite{Light-induced, Udina_2019}. While the nonlinear current and superconducting gap may exhibit similar dynamics, they originate from distinct physical mechanisms. It has been argued that in clean superconductors the nonlinear current is dominated by quasiparticle contributions rather than by the Higgs mode \cite{Cea_nonlinear}. However, subsequent studies have shown that disorder can significantly enhance the contribution of the Higgs mode to the nonlinear current response \cite{Silaev_2019,Murotani_2019,Tsuji_2020,Seibold_2021}.
% and the quasiparitlces contribution becomes less relavant in the dirty limit. 
While the Higgs mode cannot be directly observed in optical measurements on clean superconductors, we note that the gap dynamics can instead be accessed via time-resolved angle-resolved photoemission spectroscopy (tr-ARPES) \cite{smallwood_tracking_2012, Cortes_2011}. Although experimentally more challenging, probing the gap dynamics remains of considerable interest from both experimental and theoretical perspectives \cite{Schwarz_2020}.

In this manuscript, we examine the non-linear response of a clean superconductor to an optical pulse, using the Anderson pseudospin method. Our analysis goes beyond \cite{Impact_of_damping} in that (i) we allow for the possibility that the phenomenological damping rates could themselves be functions of energy (or momentum)  (ii) we consider third harmonic generation (non-linear current) in addition to gap function dynamics (iii) we consider d-wave as well as s-wave superconductors, and (iv) we consider polarization control viz. altering the polarization of the original light pulse to selectively excite and probe excitations in different irreducible representations of the underlying lattice. Our results thus significantly expand our understanding on how one may probe intrinsic relaxation in superconductors using nonlinear spectroscopy. 

This manuscript is structured as follow: In section \ref{sec: theory} we provide a pedagogical introduction to the Anderson pseudospin formalism as applied to s and d-wave superconductors. We discuss non-linear response within the Anderson pseudospin formalism, introduce (energy dependent) phenomenological damping in the equations of motion for the pseudospins, and discuss how this manifests in experimentally measurable quantities. The contents of this section are largely standard, and may be skipped by experts in the field. In Section \ref{sec: s-wave} we apply the above formalism to s-wave superconductors. Some of our results in this section were anticipated already in \cite{Impact_of_damping}, but the results on energy dependent damping, role of polarization, and third harmonic generation we believe are new. Finally, in Section \ref{sec: d-wave} we apply our formalism to d-wave superconductors (where polarization control comes into its own). We believe the results in this section to be wholly new. We conclude in Section \ref{sec: conclusions} with a 
discussion of some interesting open directions. 
\section{Theoretical background}

This section is largely pedagogical and provides theoretical background so as to make our discussion self contained. It may be skipped by experts in superconductivity. 

\label{sec: theory}

\subsection{Dynamics of Anderson pseudospins}
The microscopic theory of the superconductors can be reformulated in the basis of Nambu spinors $\psi_{k}=(c_{\boldsymbol{k}\uparrow}, c^\dagger_{-\boldsymbol{k}\downarrow})^T$, and the mean-field BCS Hamiltonian can be written as $ H = \sum_{\boldsymbol{k}} \psi_{\boldsymbol{k}}^\dagger 
\left( \boldsymbol{b}_{\boldsymbol{k}} \cdot \boldsymbol{\tau} \right) \psi_{\boldsymbol{k}} 
+$ constant, where $ \boldsymbol{\tau}$ represents the Pauli matrices. This formulation suggests that superconductivity can be mapped to a spin system by introducing the Anderson pseudospin as 
\begin{equation}
    \boldsymbol{s}_{\boldsymbol{k}}=  \frac{1}{2}\psi_{\boldsymbol{k}}^\dagger \boldsymbol{\tau} \psi_{\boldsymbol{k}}
\end{equation}
The z-component $s^z_{\mathbf{k}}$ represents the momentum distribution of electrons, while the x-
and y-components correspond to the real and imaginary parts of the Cooper pair density, respectively. 
\begin{align}
{s}^x_{\mathbf{k}} &= \frac{1}{2}
\left( c_{\mathbf{k}\uparrow}^\dagger c_{-\mathbf{k}\downarrow}^\dagger + c_{-\mathbf{k}\downarrow} c_{\mathbf{k}\uparrow} \right); \nonumber\\
{s}^y_{\mathbf{k}}& =
\frac{-i}{2} \left( c_{\mathbf{k}\uparrow}^\dagger c_{-\mathbf{k}\downarrow}^\dagger - c_{-\mathbf{k}\downarrow} c_{\mathbf{k}\uparrow} \right); \\
s^z_{\mathbf{k}} &=
\frac{1}{2}  \left( c_{\mathbf{k}\uparrow}^\dagger c_{\mathbf{k}\uparrow} - c_{-\mathbf{k}\downarrow} c_{-\mathbf{k}\downarrow}^\dagger \right) 
= \frac{1}{2} \left( n_{\mathbf{k}\uparrow}+  n_{\mathbf{-k}\downarrow}-1\right ) \nonumber
\end{align}
Up to a constant, the BCS Hamiltonian reduces to:
\begin{equation}
 H=2 \sum_{\boldsymbol{k}} \boldsymbol{b}_{\boldsymbol{k}} \cdot\boldsymbol{s}_{\boldsymbol{k}}
\end{equation}
where the pseudomagnetic field $\mathbf{b}_{\boldsymbol{k}} = 
(-\Delta', -\Delta'', \tilde{\varepsilon}_{\boldsymbol{k}}-\mu)  
 $. Here, $\tilde{\varepsilon}_{\boldsymbol{k}}-\mu=\varepsilon_{\boldsymbol{k}}$ is the energy dispersion measured from the Fermi level with $\tilde{\varepsilon}_{\boldsymbol{k}}$ and $\mu $ the single-particle band structure and the chemical potential, respectively. 
% = -\mathbf{h}_{\mathbf{k}}$.  
The complex superconducting order parameter is defined as $\Delta=\Delta'+i\Delta''$. In the ground state of s-wave superconductors at zero temperature, the pseudospins form a unique texture $(s^x_{\boldsymbol{k}},s^y_{\boldsymbol{k}},s^z_{\boldsymbol{k}})=(\Delta_0,0,-\varepsilon_{\boldsymbol{k}})/\omega_{\boldsymbol{k}}$ with natural frequency $\omega_{\boldsymbol{k}}=2\sqrt{\varepsilon^2_{\boldsymbol{k}}+\Delta_0^2}$, where the order parameter is taken real without loss of generality. In the nonequilibrium regime, the superconducting state is governed by the pseudospin dynamics, which obeys a Bloch-like equation of motion: 
\begin{equation}
    \frac{\partial}{\partial t} \boldsymbol{s}_{\boldsymbol{k}} = i[H,\boldsymbol{s_k}]=2 \boldsymbol{b}_{\boldsymbol{k}} \times \boldsymbol{s}_{\boldsymbol{k}}.
    \label{differential}
\end{equation}
This equation describes the collective precession of pseudsopins around the pseudomagnetic fields. The complex order parameter evolves as: 
\begin{equation}
    \Delta(t) = V_0 \sum_{\boldsymbol{k}} \left[s_{\boldsymbol{k}}^x(t) + i s_{\boldsymbol{k}}^y(t)\right],
    \label{self-consistent}
\end{equation}
Since the motions of pseudospins induce a temporal variation in the superconducting order parameter, this, in turn, modifies the \( x \)- and \( y \)-components of the pseudomagnetic field. As a result,  Eq. \ref{differential} and Eq. \ref{self-consistent} need to be solved self-consistently. This Anderson pseudospin formalism provides a clear connection between pseudospin dynamics and collective excitations in superconductors.

For d-wave superconductors, the anisotropic pairing leads to a momentum-dependent energy gap $\Delta_{\boldsymbol{k}}=\frac{1}{2}\Delta_0(\cos k_x-\cos k_y) $. There are nodal lines at $\phi_{\boldsymbol{k}}==\pm 45^\circ$. The corresponding equilibrium pseudospin texture is $\boldsymbol{s}_{\boldsymbol{k}}
= \left(
\frac{\Delta_{\boldsymbol{k}}}{\omega_{\boldsymbol{k}}},
\, 0,\,
-\frac{\varepsilon_{\boldsymbol{k}}}{\omega_{\boldsymbol{k}}}
\right)$ where $\omega_{\boldsymbol{k}} = 2\sqrt{\varepsilon_{\boldsymbol{k}}^2 + \Delta_{\boldsymbol{k}}^2}$. 
The gap dynamics is governed by the pseudospin evolution as  $\Delta_{\boldsymbol{k}}(t)=\sum_{\boldsymbol{k'}} V_{\boldsymbol{k}\boldsymbol{k'}}(s^x_{\boldsymbol{k'}}(t)+is^y_{\boldsymbol{k'}}(t))$ where
\begin{equation}
    V_{\boldsymbol{k}\boldsymbol{k'}}= \frac{V_0}{4} (\cos k_x-\cos k_y)(\cos k'_x-\cos k'_y)
\end{equation}
It follows that 
\begin{equation}
\Delta_{\boldsymbol{k}}(t)=\frac{1}{2}{\Delta}(t)(\cos k_x-\cos k_y)
\end{equation}
where ${\Delta}(t)=\frac{1}{2} V_0\sum_{k'}  (\cos k'_x-\cos k'_y)(s^x_{k'}(t)+ i s^y_{k'}(t))$. With this choice of d-wave pairing, the nonequilibrium gap dynamics retains the same momentum dependence as the equilibrium gap function, with its overall magnitude  $|{\Delta}(t)|$ evolving in time.  

\subsection{Nonlinear light coupling and polarization effect}

In an \textit{s}-wave superconductor, the optical gap edge is located at  $2\Delta$. Therefore, high-frequency excitations with $\omega > 2 \Delta$ break Cooper pairs into quasiparticles via linear coupling.  However, the Higgs mode, being a scalar excitation, does not couple linearly to electromagnetic fields, making it challenging  to observe the Higgs mode through light. Recent experimental studies have shown that lower-frequency radiation in the terahertz (THz) range, with a center frequency $\omega  \approx\Delta$,  can couple nonlinearly to the superconducting 
state and induce oscillation of the order parameter \cite{Matsunaga, Light-induced}. 

We model the light via minimal coupling in the particle and hole sectors, which carry opposite charges. This yields the pseudomagnetic field
$
\boldsymbol{b}_{\boldsymbol{k}}(t) = 
\begin{pmatrix}
-\Delta'(t), -\Delta''(t), \frac{\tilde\varepsilon_{\boldsymbol{k}-\frac{e}{c}\boldsymbol{A}(t)} + \tilde\varepsilon_{\boldsymbol{k}+\frac{e}{c}\boldsymbol{A}(t)}}{2} -\mu
\end{pmatrix}.$ We also assume that the system is parity symmetric, $\tilde\varepsilon_{\boldsymbol{k}}=\tilde\varepsilon_{\boldsymbol{-k}}$, so the odd powers in $A$ cancel and the leading light–matter coupling is quadratic: 
\begin{equation}
    \boldsymbol{b}^z_{\boldsymbol{k}}=
    {\tilde\varepsilon}_{\boldsymbol{k}} -\mu + \frac{e^2}{2c^2} \sum_{i,j} \frac{\partial^2 \tilde\varepsilon_{\boldsymbol{k}}}{\partial k_i \partial  k_j} A_i(t) A_j(t) .
 \end{equation} 
Throughout this work, we focus on the long-wavelength limit with $q=0$. In this regime the vector potential can be treated as spatially uniform. 
% The light pulse drives the superconductor out of equilibrium and induces a coherent collective precession of the pseudospins. 

We note that for an isotropic parabolic band,  $\partial_{k_i} \partial_{k_j}\tilde \epsilon_{\boldsymbol{k}}=(\hbar^2 /m^* )\delta_{ij}$, the light-induced term
% change $\delta b^z_k(t)=\frac{e^2}{2m^*c^2}|A(t)|^2$ 
becomes independent of the crystal momentum for any light polarization. Therefore, it can be interpreted as a time-dependent chemical potential shift and does not vary the amplitude of the order parameter. 
It turns out to be essential to adopt a dispersion with $k$-dependent curvature to observe nontrivial dynamics. Let's work on a 2D square lattice with a band dispersion $\tilde\varepsilon_k=-2J(\cos k_xa+\cos k_ya) $ and consider a linear polarized light $A_0(t)(\cos\alpha,\sin\alpha,0)$ where $\alpha$ is the  light polarization. The variation of the pseudomagnetic field can be expressed as
\begin{equation}
    \delta b^z_k(t) =  \frac{e^2}{2c^2} \sum_{i,j} \frac{\partial^2 \tilde\varepsilon_{\mathbf{k}}}{\partial k_i \partial k_j} A_i(t) A_j(t) = {\tilde A}^2(t)F_k(\alpha)
    \label{F_k}
\end{equation}
where $\tilde A^2(t)=Ja^2\frac{e^2}{c^2} A^2_0(t)$. The tensor $\frac{\partial^2 \tilde\varepsilon_{\mathbf{k}}} {\partial k_i \partial k_j}$ is analogous to the Raman tensor, but note that both $A_i$ and $A_j$ represent incoming fields in this case. The function $F_k$ encodes the information of the light polarization. 
\begin{align}
    F_k (\alpha)=\cos^2 \alpha \cos k_x +\sin^2 \alpha\cos k_y
\end{align}
The irreducible representations of the $D_{4h}$ point group in two dimensions can be expressed as
\begin{align}
    A_{1g} &= \cos{k_x}+\cos{k_y} \nonumber\\
    B_{1g} &= \cos{k_x}-\cos{k_y} \\ 
    B_{2g} &= \sin k_x  \sin k_y  \nonumber
\end{align}
For $x$-polarized light, $\alpha=0$, 
\begin{align}
F_k(0) &= \cos k_x  \notag =\tfrac{1}{2}\Big[(\cos k_x + \cos k_y) + (\cos k_x - \cos k_y)\Big] \nonumber\\
&=\left(C_0-\frac{\varepsilon_{k}}{4J}\right)+  \frac{1}{2}(\cos k_x - \cos k_y)\in A_{1g}+B_{1g}
\label{const}
\end{align}
 where $C_0=-\mu/4J$ is a constant ($k$-independent) term and is irrelavant to the dynamics. 
 
 For $x'$-polarized light (along 45 $^\circ$ direction), $\alpha=\pi/4 $, and
\begin{equation}
F_k(\pi/4)=\cos{k_x}+\cos{k_y}  \in A_{1g}.
\end{equation}
Thus, rotating the light polarization selects different mixtures of irreps. Polarization can accordingly be used to selectively probe modes in different irreps - as was done experimentally in \cite{Katsumietal}. 

Suppose we consider the second and third-nearest-neighbor hopping $J'$ and $J''$ in a 2D square lattice, giving rise to the energy dispersion $ \tilde\varepsilon_k= -2J(\cos{k_xa}+\cos{k_ya})-4J'\cos{k_xa} \cos{k_ya} - 2J''(\cos{2k_xa}+\cos{2k_ya})$. Setting $a=1$, we obtain

\begin{widetext}
\begin{equation}
\frac{\partial^2 \tilde\varepsilon_{\mathbf{k}}}{\partial k_i \partial k_j} =
\begin{pmatrix}
2J \cos k_x + 4J' \cos k_x \cos k_y + 8J'' \cos 2k_x & -4J' \sin k_x \sin k_y \\
-4J' \sin k_x \sin k_y & 2J \cos k_y + 4J' \cos k_x \cos k_y + 8J'' \cos 2k_y
\end{pmatrix}
\label{tensor}
\end{equation}

\begin{align}
    F_k(\alpha)=& 2J \left( \cos^2\alpha \cos k_x  +\sin^2\alpha \cos k_y \right) + 8J''\left( \cos^2\alpha \cos 2k_x  +\sin^2\alpha \cos 2k_y \right) \nonumber \\
    +& 4J' \cos k_x \cos k_y -8J' \sin \alpha \cos \alpha \sin k_x \sin k_y
\end{align}

\end{widetext}
For x-polarized light,  
\begin{equation}
    F_k(\alpha=0) \in A_{1g} +B_{1g}
    \label{F_k_x}
\end{equation}
For x'-polarized light, 
\begin{equation}
    F_k(\alpha=\pi/4) \in A_{1g} +B_{2g}
    \label{F_k_x_prime}
\end{equation}
The function $F_k (\alpha)$ determines the change in the pseudomagnetic field during the light pulse, and the pseudospin texture is correspondingly twisted.  

\subsection{Pseudospin texture \label{Pseudospin_texture}}
Understanding the nonequilibrium pseudospin texture is crucial for analyzing the gap dynamics. 
We consider the initial condition of pseudospins $s_k(0)=(\Delta_k/\omega_k,0, -\varepsilon_k/\omega_k )  $ where $\omega_k=2\sqrt{\varepsilon_k^2+\Delta_k^2}$. Within a linear analysis, we assume the pseudospins $\boldsymbol{s_k}(t)=\boldsymbol{s_k}(0)+\delta \boldsymbol{s_k}(t)$ and $\Delta_k(t)=\Delta^{0}_k+\delta \Delta_k(t)$. One may wonder why it is acceptable to ignore phase fluctuations of the order parameter. This is because the $k$-independent light-induced term (e.g. $C_0$ in Eq. \ref{const}) only contributes a trivial global phase $e^{4iC_0\int_0^t\tilde A^2(t')dt'}$ to $\Delta(t)$ and can be gauged out. Therefore, we only need to consider the $k$-dependent light-induced term in the following analysis. Moreover, for a narrow Debye shell compared to the band width ($\Delta/J \ll1 $), the DOS is approximately particle–hole symmetric near the Fermi surface. Therefore, we can assume that the Higgs and phase mode are effectively decoupled. 
% We note that for an constant,  As a result, the order parameter $\Delta_k(t)=\Delta^{0}_k+\delta \Delta_k(t)$ can be chosen real.
% with pseudomagnetic field is $\mathbf{b}_{\mathbf{k}}(t) = 
% \begin{pmatrix}
% -\Delta_k-\delta\Delta(t), 0, \varepsilon_k+\frac{e^2}{2c^2} \sum_{ij} \partial_{k_i} \partial_{k_j} \epsilon_{\boldsymbol{k}} A_i(t) A_j(t)\end{pmatrix}$. 
The linearized equations of motion for the pseudospins in the co-rotating frame are given as \cite{Theory_of_Anderson}:
\begin{align}
    \partial_t \delta s_{\boldsymbol{k}}^x (t) &= -2 \epsilon_{\boldsymbol{k}} \delta s_{\boldsymbol{k}}^y (t), \nonumber \\
    \partial_t \delta s_{\boldsymbol{k}}^y (t) &= 2 \epsilon_{\boldsymbol{k}} \delta s_{\boldsymbol{k}}^x (t) +2\Delta \, \delta s_{\boldsymbol{k}}^z (t) + \frac{1}{2 \epsilon_{\boldsymbol{k}}} g_k(t)\\
    \partial_t \delta s_{\boldsymbol{k}}^z (t) &= -2\Delta \, \delta \nonumber s_{\boldsymbol{k}}^y (t).
\end{align}
where 
\begin{equation}
    g_k(t) = \frac{4 \epsilon_{\boldsymbol{k}}\Delta_k}{\omega_{\boldsymbol{k}}} \left[ \frac{e^2}{2c^2}  \sum_{ij} \frac{\partial^2 \varepsilon_{\mathbf{k}}}{\partial k_i \partial k_j}A_i (t) A_j (t) -  \varepsilon_{\boldsymbol{k}} \frac{\delta \Delta_k (t)}{\Delta_k} \right]
    \label{driving}
\end{equation}
From the first and third equations, it can be seen that $\Delta_k \partial_t \delta s_{\boldsymbol{k}}^x (t) =  \varepsilon_k \partial_t \delta s_{\boldsymbol{k}}^z (t)$. Since the initial condition is $\delta\boldsymbol{s_k}(0)=0$, we have $\Delta \delta s_{\boldsymbol{k}}^x (t) =\varepsilon_k\delta s_{\boldsymbol{k}}^z (t) $ for all times. By collecting the terms involving
$\delta s^x_k(t)$, we arrive at $\partial^2_t \delta s_{\boldsymbol{k}}^x(t) = -\omega^2_k \delta s_{\boldsymbol{k}}^x(t)-g_k(t) $. Expressed in the Fourier space:
\begin{equation}
    s^x_k(\omega)=\frac{g_k(\omega)}{\omega^2-\omega^2_k}
\end{equation}
By plugging Eq.\;\ref{driving} and combining with Eq.\;\ref{F_k}, we obtain
\begin{align}
\delta s^x_k(\omega) &= \frac{4\varepsilon_k\Delta_k}{\omega_k} \frac{1}{\omega^2-\omega_k^2}\left[A^2(\omega) F_k(\alpha)-\varepsilon_k  \frac{\delta\Delta_k(\omega)}{\Delta_k} \right] \nonumber \\
\delta s^y_k(\omega) &= -i\frac{2\Delta_k}{\omega_k} \frac{\omega}{\omega^2-\omega_k^2}\left[A^2(\omega) F_k(\alpha)-\varepsilon_k  \frac{\delta\Delta_k(\omega)}{\Delta_k} \right] \nonumber \\
\delta s^z_k(\omega) &= \frac{4\Delta^2_k}{\omega_k} \frac{1}{\omega^2-\omega_k^2}\left[A^2(\omega) F_k(\alpha)-\varepsilon_k  \frac{\delta\Delta_k(\omega)}{\Delta_k} \right]
\end{align}  where the x,y and z-components are related by $\delta s^z_k(\omega)=(\Delta_k/\epsilon_k) \delta s^x_k(\omega)$ and $\delta s^y_k(\omega)= -\frac{i\omega}{2\epsilon_k} \delta s^x_k(\omega)$. With the inverse Fourier transformation $F^{-1}[\frac{1}{\omega^2-\omega^2_k}]=\frac{\sin\omega_k t}{\omega_k} \Theta(t)$, the pseudospin twisting dynamics can be expressed as a convolution integral:
\begin{align}
    \delta s^x_k(t) &=\frac{4\epsilon_k\Delta_k}{\omega^2_k} \int_{0}^{t} d\tau  \sin{\omega_k(t-\tau)} G_{\alpha, k} (\tau) \nonumber \\ 
     \delta s^z_k(t) &=\frac{4\Delta^2_k}{\omega^2_k} \int_{0}^{t} d\tau  \sin{\omega_k(t-\tau)} G_{\alpha, k}(\tau) \label{22} \\  
     G_{\alpha, k}(\tau) &= A^2(\tau)F_k(\alpha)-\varepsilon_k \frac{\delta\Delta_k(\tau)}{\Delta_k} \nonumber
\end{align}

An important observation is that the pseudospin twisting is determined by the competition between the direct light-induced term $A^2(\tau)F_k(\alpha)$ and the self-gap dynamics term $\varepsilon_k \frac{\delta\Delta_k(\tau)}{\Delta_k}$. Using Eqs.\;\ref{F_k_x},\ref{F_k_x_prime} for function $F_k(\alpha)$ and the fact that the gap term $\varepsilon_k\frac{\delta\Delta_k(\tau)}{\Delta_k} \in A_{1g}$, we can analyze the induced irreps of the superconducting condensate. For $d$-wave superconductors, the $x$-component of the psuedospin has a prefactor $\Delta_k\in B_{1g}$. Consequently, $\delta s_k^x \in B_{1g}\times(A_{1g}+B_{1g}) =B_{1g}+A_{1g} $ for $x$-polarized light and $\delta s_k^x \in  B_{1g}\times(A_{1g}+B_{2g}) =B_{1g}+A_{2g} $ for $x$'-polarized light. On the other hand, the prefactor of the $z$-component is $\Delta_k^2 \in A_{1g}$. Therefore, $\delta s_k^z \in A_{1g}+B_{1g}$ for $x$-polairzed light and $\delta s_k^z \in  A_{1g}+B_{2g}$ for $x$'-polaized light. Interestingly, Eq.\;\ref{22} also implies that the irreps of the pseudospin texture can acquire nontrivial energy dependence. 
Taking $\delta s_k^z$ as an example, the contribution arising from the gap term contains only an $A_{1g}$ component, whose magnitude grows with increasing energy $\varepsilon_k$. In contrast, the light term can induce different irreducible representations; it dominates near the Fermi surface but decays away from it if the light term $A^2(\omega)$ is a Gaussian function centered at $\omega=0$. As a result, if the light term induces a $B_{1g}$ component near the Fermi surface, it decays with increasing $\varepsilon$, and eventually the $A_{1g}$ component dominates far from the Fermi surface. 

In this work, we will mostly focus on states near the Fermi surface, because we are interested in the long-time behavior of the current and gap dynamics, which is dominated by low-energy states, as pseudospins with higher energy gradually dephase. In addition, once pseudospin relaxation is introduced, it is natural to expect that the relaxation rate grows with energy, so that high-energy pseudospins decay much faster. 
Finally, we comment on the pseudospin twisting in the impulsive limit with $\tilde A(t) =  A\,\delta(t)$ and assuming that the gap is suddenly quenched, $\delta\Delta_k(t) = \delta\Delta_k \,\Theta(t)$ (note that the gap is oscillatory, but here we neglect the oscillation for simplicity). In this case, the $x$-component can be expressed as

\begin{align}
    \delta s^x_k(t) &=\frac{4\epsilon_k\Delta_k}{\omega^2_k} \int_{0}^{t} d\tau  \sin{\omega_k(t-\tau)}  \left[A^2 \delta(\tau)f_k(\alpha) -\varepsilon_k \frac{\delta\Delta_k}{\Delta_k}\right]  \nonumber \\
    &=\frac{4\epsilon_k\Delta_k}{\omega^2_k}   \left[A^2 f_k(\alpha) \sin{\omega_k t} +\delta \Delta_k\frac{ \varepsilon_k}{\omega_k}(\cos \omega_kt-1)\right] 
\end{align}

\subsection{Nonlinear current for different irreducible representations \label{irreps}}

While the energy gap only probes pseudospins with irreps corresponding to the gap symmetry, the nonlinear current allows us to probe additional irreps. The nonlinear current can be expressed in the pseudospin formalism by summing over the variation of the $z$-component, and is directly related to the third-harmonic generation (THG) \cite{Cea_nonlinear, Puviani_quench}
\begin{align}
    J_i^{(3)}(t) &= {-2e^2}\sum_{k j} A_j(t) 
    \frac{\partial^2 \varepsilon_{\mathbf{k}}}{\partial k_i \partial k_j} 
    \,\delta s^z_k(t) 
    % \\
    % &= \sum_{k j} A_j(t) \frac{\partial^2 \varepsilon_{\mathbf{k}}}{\partial k_i \partial k_j} 
    %    \sum_{l,m} \frac{\partial^2 \varepsilon_{\mathbf{k}}}{\partial k_l \partial k_m}  
    %    A_l(t) A_m(t)    
\end{align}

In this work, we consider a pump–probe setup in which pseudospins are twisted by a strong pump pulse at $t=0$, inducing pseudospin precession, and subsequently probed at a delay time $t=\tau$ by a short and weak probe pulse (whose effect on the pseudospins can be neglected, and the probe pulse does not overlap with the pump pulse).  For a band dispersion including second- and third-nearest-neighbor hoppings, the tensor structure of Eq.\;\ref{tensor} can be used to analyze the irreps. If both the probe pulse and the current are measured along the $x$ direction, then the tensor contains $A_{1g}$ and $B_{1g}$ components, implying that both $A_{1g}$ and $B_{1g}$ components of $s^z_k$ are probed. If both the probe pulse and the current are measured along the $x'$ direction (45$^\circ$), then $A_{1g}+B_{2g}$ pseudospins are probed. If the current is measured in a direction perpendicular to the probe polarization, only the $B_{2g}$ channel is accessed.  

To summarize, we introduce the notation $J_{ijkl}$, where $i$ is the direction of the measured current, $j$ is the polarization of the weak probe pulse, and $(k,l)$ are the polarizations of the strong quench pulse that twists the pseudospins. For $x$-polarized quench pulse we obtain
\begin{align}
    J_{xxxx}   &= J_{A_{1g}+B_{1g}}, \nonumber \\ 
    J_{x'x'xx} &= J_{A_{1g}+B_{2g}} = J_{A_{1g}}, \\
    J_{xyxx}   &= J_{yxxx} = J_{B_{2g}} = 0 \nonumber,
\end{align}
while for $x'$-polarized quench light we obtain
\begin{align}
    J_{xxx'x'}   &= J_{A_{1g}+B_{1g}}, \nonumber\\ 
    J_{x'x'x'x'} &= J_{A_{1g}+B_{2g}}, \\
    J_{xyx'x'}   &= J_{yxx'x'} = J_{B_{2g}} \ne 0 \nonumber.
\end{align}

The $J_X$ notation with $X=A_{1g},B_{1g},B_{2g} $ means that the $X$ irreps of the ``pseudospin $z$- component" can be probed from the nonlinear current. The dynamics of the nonlinear current can be measured by tuning the delay time of the probe pulse (e.g., using a short and weak Gaussian pulse). Since the pseudospin $z$-component undergoes precession, the nonlinear current will also oscillate, analogous to the Higgs oscillation of the gap. Consequently, we expect the oscillation amplitude of the nonlinear current to decay even in the collisionless limit, similar to the Higgs mode decay. In both cases, the intrinsic decay of the macroscopic signal can be attributed to microscopic pseudospin dephasing.

\subsection{Pseudospin relaxation times $T_1$ and $T_2$}

So far, we have described the pseudospins in Cartesian coordinates and assumed that there are no decay channels. Alternatively, we can decompose the pseudospin as
\begin{equation}
    \vec{s}_k(t) = s_k^\parallel(t) + s_k^\perp(t)
\end{equation}
where $s_k^\parallel(t)$ and $s_k^\perp(t)$ denote the longitudinal and transverse components with respect to the pseudomagnetic field $\boldsymbol{b}_k(t)$. The longitudinal component ${s}_k^{\parallel}$, which is the projection of each pseudospin onto its pseudomagnetic field, is related to the Bogoliubov quasiparticle occupation \( n^{qp}_{k} \) via
\begin{equation}
{s}_{{k}}^{\parallel} ={s}^0_{{k}}\,(1 - 2 n^{qp}_{{k}}),
\end{equation}
where \( {s}^0_{{k}} \) is the equilibrium pseudospin at $T=0$. When $n^{qp}_{{k}}=1$, the pseudospin is flipped, corresponding to the creation of a pair of Bogoliubov quasiparticles at momentum states \( \mathbf{k}\uparrow \) and \( -\mathbf{k}\downarrow \). The creation of a single quasiparticle is not allowed in the Anderson pseudospin formalism, which only applies within the mean-field approximation (i.e., pseudospins are not removed). At finite temperature, the quasiparticle number follows the Fermi–Dirac distribution,
$
n^{qp}_\mathbf{k} = \frac{1}{e^{\beta E_\mathbf{k}} + 1}.
$
Consequently, the pseudospin texture becomes
$
\boldsymbol{s}_\mathbf{k}(T) = \boldsymbol{s}^0_\mathbf{k} \tanh\!\left(\frac{\beta E_\mathbf{k}}{2}\right),
$
which shows that the longitudinal component is reduced at finite temperature.

The pseudospin precession around its pseudomagnetic field at the natural frequency $\omega_k$ can be described by Bloch equations, analogous to the dynamics of a two-level system. It is therefore natural to characterize pseudospin relaxation using the standard relaxation times $T_1$ and $T_2$. The longitudinal component $s^\parallel_k$ is interpreted as the quasiparticle distribution; its relaxation is governed by $T_1$, the quasiparticle redistribution time. The transverse component $s^\perp_k$ characterizes pseudospin coherence; its relaxation is governed by $T_2$, the quasiparticle dephasing time. Thus, the dynamics of a pseudospin can be approximated as
\begin{align}
    \delta s^\perp_k(t) &= \delta s^\perp_k(0^+) \,\cos(\omega_k t)\, e^{-t/T_{2,k}}, \\
    \delta s^\parallel_k(t) & 
      = \delta s^\parallel_k(0^+) e^{-t/T_{1,k}}
    % \delta s^\parallel_k(t) &= \tfrac{1}{2}-s^\parallel_k(t)
    %   = \left[\tfrac{1}{2}-s^\parallel_k(0^+)\right] e^{-t/T_{1,k}},
\end{align}
where $\delta s^\perp_k(0^+)$ and $\delta s^\parallel_k(0^+)$ denote the pseudospin twisting immediately after the pump pulse at $t=0$. Note that the relaxation times $T_{1,k}$ and $T_{2,k}$ may depend on the momentum $\mathbf{k}$.

The Cartesian pseudospin components $s^x_k, s^y_k, s^z_k$ are linear combinations of the longitudinal and transverse components. In the long-time limit when relaxation is included, the pseudospin precesses while remaining roughly pointing along $ (\Delta_k,0,-\varepsilon_k)$. The $x$-component can thus be expressed as
\begin{equation}
    s^x_k(t) =\frac{\Delta_k}{E_k} s^\parallel_k(t) - \frac{\varepsilon_k}{E_k} s^\perp_k(t)
\end{equation}
and a similar relation holds for the $z$-component. As a result, macroscopic observables such as the superconducting gap and the nonlinear current contain contributions from both $T_1$ and $T_2$. For instance, in $s$-wave superconductors, the gap takes the form
\begin{align}
\Delta(t) &= \Delta_0
+ \sum_k \frac{\Delta_k}{E_k}\,\delta s^\parallel_k(0^+)\, e^{-t/T_{1,k}} \nonumber \\
& - \sum_k \frac{\varepsilon_k}{E_k}\,\delta s^\perp_k(0^+)\, e^{-t/T_{2,k}} \cos(\omega_k t) \nonumber  \\
 &= \Delta_0 - \Delta^R(t) + \Delta^O(t),
\end{align}
where $\Delta^R(t)$ and $\Delta^O(t)$ represent the recovery and oscillatory components of the gap, respectively. Thus, by separating the decay of the recovery and oscillatory parts of macroscopic observables, one can in principle extract the microscopic pseudospin lifetimes $T_{1,k}$ and $T_{2,k}$.

In this work, we consider three types of damping: (i) homogeneous damping with $1/T_k=\text{const.}$, (ii) inhomogeneous damping with $1/T_k \propto \omega_k^p$, and (iii) inhomogeneous damping with $1/T_k \propto |\varepsilon_k|^p$. In the presence of inhomogeneous damping, one may ask whether the effective relaxation times should be averaged over the Debye shell, and whether such inhomogeneity induces additional broadening. We will discuss these issues and explain how $T_{1,k}$ and $T_{2,k}$ can be extracted from macroscopic observables in both $s$- and $d$-wave superconductors.

\section{$s$-wave superconductors}
\label{sec: s-wave} 

We consider an $s$-wave superconductor on a 2D square lattice with a single-particle band dispersion 
$\tilde{\varepsilon}_\mathbf{k} = -2J\left(\cos k_x a + \cos k_y a \right),$
where $J$ is the nearest-neighbor hopping amplitude in the tight-binding model, chosen to be much greater than the superconducting gap ($J/\Delta = 10^3$). We choose $\mu = -3J$ so that the Fermi surface is approximately circular (and well away from Van Hove filling). We model the pump pulse using a Gaussian vector potential 
$\boldsymbol{ A}(t) =  A_0 \exp\!\left[-\tfrac{8 \Delta_0^2 (t - \pi / \Delta_0)^2}{\pi^2}\right]\hat{n}$, 
where $\hat{n}$ is the unit polarization vector and $A_0$ is the peak amplitude of the vector potential. The corresponding electric field is given by 
$\boldsymbol{E}(t) = -\tfrac{1}{c}\,\tfrac{d\boldsymbol{A}(t)}{dt}$, 
which has the form of the derivative of a Gaussian envelope and thus represents a single-cycle pump pulse. 

\begin{figure*}[htb]
    \centering
    \includegraphics[width=0.85\textwidth]{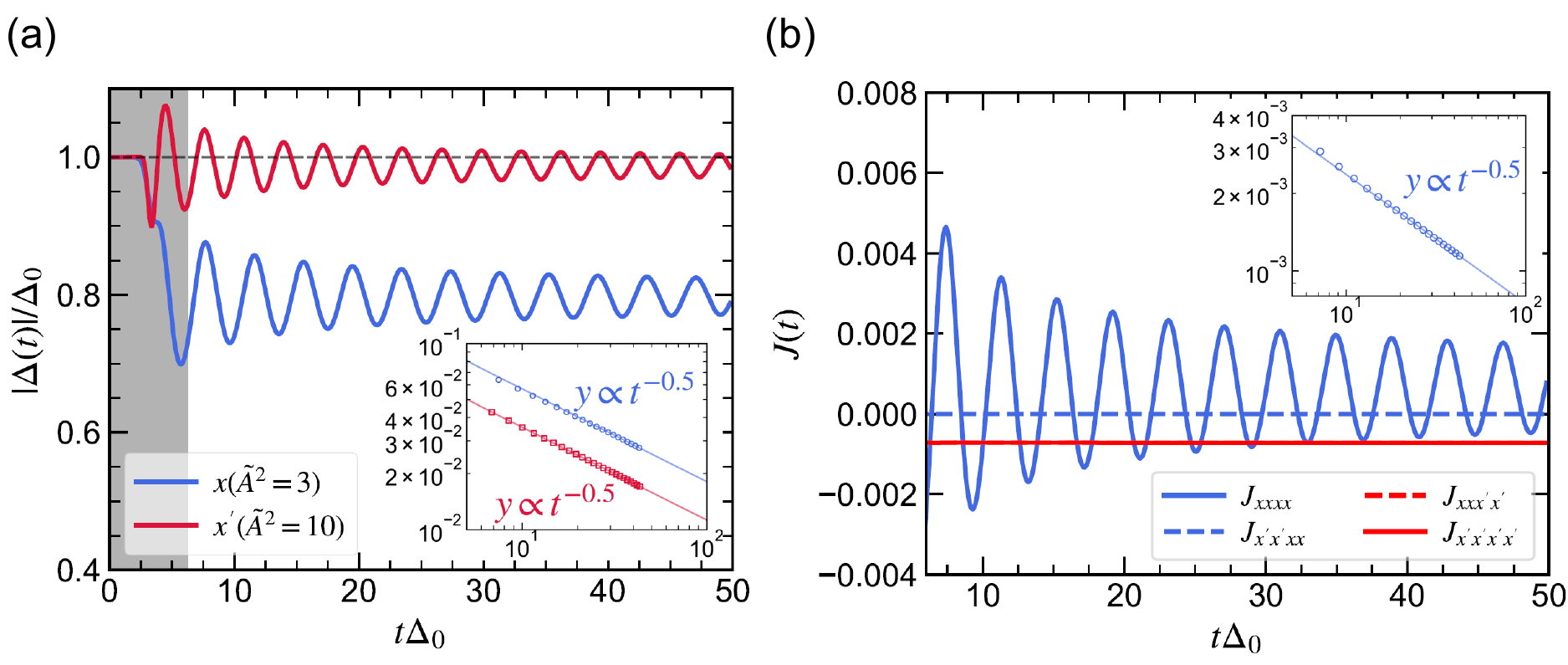}
    \caption{ (a) Gap dynamics in the collisionless limit ($1/T_1=1/T_2=0$) with $x$- and $x'$-polarized pump pulses. The shaded area marks the duration of the light pulse. Inset: Higgs amplitude decay on a log–log scale, showing a $t^{-1/2}$ power-law behavior. 
    (b) Nonlinear current dynamics in the collisionless limit. Inset: amplitude of $J_{xxxx}=J_{A_{1g}+B_{1g}}$ on a log–log scale, which decays as $1/\sqrt{t}$, analogous to the Higgs amplitude mode. The nonlinear current is shown in arbitrary units. }
    \label{fig:1}
\end{figure*}

\subsection{Collisionless limit}

In standard spin echo experiments, the relaxation times satisfy
\begin{align}
 \frac{1}{T_2} &= \frac{1}{2T_1} + \frac{1}{T_2^*}, \label{T1T2}\\
 \frac{1}{T'_2} &= \frac{1}{T_2} + \frac{1}{T^{\text{inhom}}_2}.
\end{align}
Here $T_1$ is the intrinsic energy relaxation timescale and $T_2$ is the intrinsic dephasing timescale. Meanwhile $T_2^*$ reflects `pure' dephasing over and above the dephasing induced by energy relaxation, and $T_2'$ is the dephasing timescale that would be measured in a coarse grained experiment (that includes the effects of inhomogenous broadening from nonuniform pseudomagnetic field, captured by $T_2^{inhom}$). %\RN{What is $T_2'$?} where $T_2^*$ is the pure dephasing time \RN{meaning what?}, and $T^{\text{inhom}}_2$ is the dephasing time caused by inhomogeneous local magnetic fields. Thus, even in the collisionless limit ($1/T_1 = 1/T_2 = 0$), the macroscopic signal still decays due to $T^{\text{inhom}}_2$. 

For Anderson pseudospins in superconductors, the inhomogeneity of the pseudomagnetic field is intrinsic: pseudospins precess around their pseudomagnetic fields with different natural frequencies, leading to a power-law decay of the amplitude oscillation instead of an exponential decay. For $s$-wave superconductors, the oscillation of macroscopic observable $O(t)$ (gap or nonlinear current) can be approximated by an integral over energy $\varepsilon$,
\begin{align}
O(t) &\sim \int_{-\infty}^{\infty} d\varepsilon \, f(\varepsilon) \cos[\omega(\varepsilon) t] \nonumber \\
&\propto f(\varepsilon=0)\, \frac{\cos(2\Delta t + \pi/4)}{\sqrt{t}}.
\end{align}
Here, $\omega(\varepsilon)$ denotes the pseudospin precession frequency, and $f(\varepsilon)$ is a smooth weighting function that is nonzero at the Fermi surface. In particular, the explicit form of $f(\varepsilon)$  for Higgs oscillations has been derived for interaction quenches \cite{Dynamical_Vanishing}. Applying the stationary-phase approximation in the long-time limit then yields the characteristic $1/\sqrt{t}$ decay. We thus expect that the intrinsic inhomogeneity leads both the Higgs amplitude mode and the nonlinear current in clean superconductors to exhibit power-law decay.

Our numerical results for the gap dynamics under $x$- and $x'$-polarized light are shown in Fig.~\ref{fig:1}(a). During the pump pulse, the energy gap is quenched, and at long times it asymptotically approaches a constant value $\Delta_\infty$, which depends on the pump intensity. We find that the gap dynamics are well described by
\begin{equation}
    \frac{|\Delta(t)|}{\Delta_\infty} 
    = 1 + a \, \frac{\cos(2\Delta_\infty t + \phi)}{\sqrt{\Delta_\infty t}},
\end{equation}
indicating that the post-pump dynamics closely resemble those of an interaction quench \cite{Dynamical_Vanishing}. The decay of the oscillation amplitude, shown in log–log scale in the inset, follows a $t^{-1/2}$ power law in the collisionless limit. This can also be attributed to Landau damping, arising from the interaction of the collective mode with the continuum of quasiparticle excitations above the gap \cite{Collisioness}. Due to this slow power-law decay, the Higgs mode is effectively considered to have an infinite lifetime. % We further note that the oscillation amplitude $a$ scales as the light intensity $A^2$, while the gap quench $\delta \Delta = \Delta_0 - \Delta_\infty$ scales as $A^4$, both indicating the nonlinear response nature. 
At sufficiently strong pumping, the gap eventually vanishes, analogous to Phase I in the interaction-quench case \cite{Dynamical_Vanishing, Fosterpip}.

Figure~\ref{fig:1}(b) shows the dynamics of the nonlinear current $J_{ijkl}(t)$. For $x$-polarized pump light, we find an oscillating signal in $J_{xxxx} = J_{A_{1g}+B_{1g}}$ but a vanishing response in $J_{x'x'xx} = J_{A_{1g}}$. Our result shows that the amplitude of $J_{xxxx}$ decays as $1/\sqrt{t}$, indicating a universal power-law behavior shared by both the nonlinear current and the Higgs oscillation.

% arise from  due to the intrinsic inhomogeneity of the pseudomagnetic field, leading to a 

\subsection{Pseudospin relaxation}

In this work, we assume that the pseudospins eventually relax toward their equilibrium values at zero temperature while preserving their spin length at all times. The damping is expressed in the following form:

\begin{equation}
    \boldsymbol{b}_{k}^{\text{eff}} = \boldsymbol{b}_{k} - \gamma_k \hat{\boldsymbol{b}}_{k} \times \boldsymbol{s}_{k}.
\end{equation}
Let $\theta_k(t)$ denote the angle between the pseudospin and its pseudomagnetic field. For this damping equation, each pseudospin evolves as $d\theta_k/dt = -\gamma_k \sin\theta_k \sim -\gamma_k\theta_k$ for small angles. Therefore, $\theta_k \propto \exp(-\gamma_k t)$ in the long-time limit. The transverse and longitudinal components can then be expressed as

\begin{align}
    |\delta s^\perp_k (t) | &= |s| \sin\theta_k \sim |s| \theta_k \propto e^{-\gamma_k t} = e^{-t/T_{2,k}}, \\
    |\delta s^\parallel_k (t)| &= |s|(1-\cos\theta_k) \sim \frac{|s|\theta_k^2}{2} \propto e^{-2\gamma_k t} = e^{-t/T_{1,k}}.
\end{align}
We note that the decay of the longitudinal pseudospin component is related to $T_1$, while the decay of the transverse pseudospin component is related to $T_2$. In our setting, 
\begin{equation}
    \gamma_k = 1/T_{2,k} = 1/(2T_{1,k})
\end{equation}
By comparing it with Eq.\;\ref{T1T2}, we conclude that no pure dephasing ($1/T^*_2=0$) is introduced in this spin-length preserving model at late times. Therefore, the $T_2$ relaxation originates entirely from $T_1$ relaxation. In other words, relaxation is modeled as energy loss to the environment, e.g., phonon degrees of freedom.  

In $s$-wave superconductors, we consider an energy-dependent relaxation rate $\gamma(\varepsilon)$ under three different models:  
(1) Homogeneous damping with $\gamma(\varepsilon)=\gamma_0\Delta_0$,  
(2) Inhomogeneous damping with $\gamma(\varepsilon)=\gamma_0 (\omega/2)^p = \gamma_0 [\varepsilon^2+\Delta_0^2]^{p/2}$, and  
(3) Inhomogeneous damping with $\gamma(\varepsilon)=\gamma_0 \Delta_0 |\varepsilon|^p$,  where $\varepsilon$ is measured from the Fermi level. In what follows, we set $\Delta_0=1$ for convenience.  

\begin{figure*}[htb]
    \centering
    \includegraphics[width=0.99\textwidth]{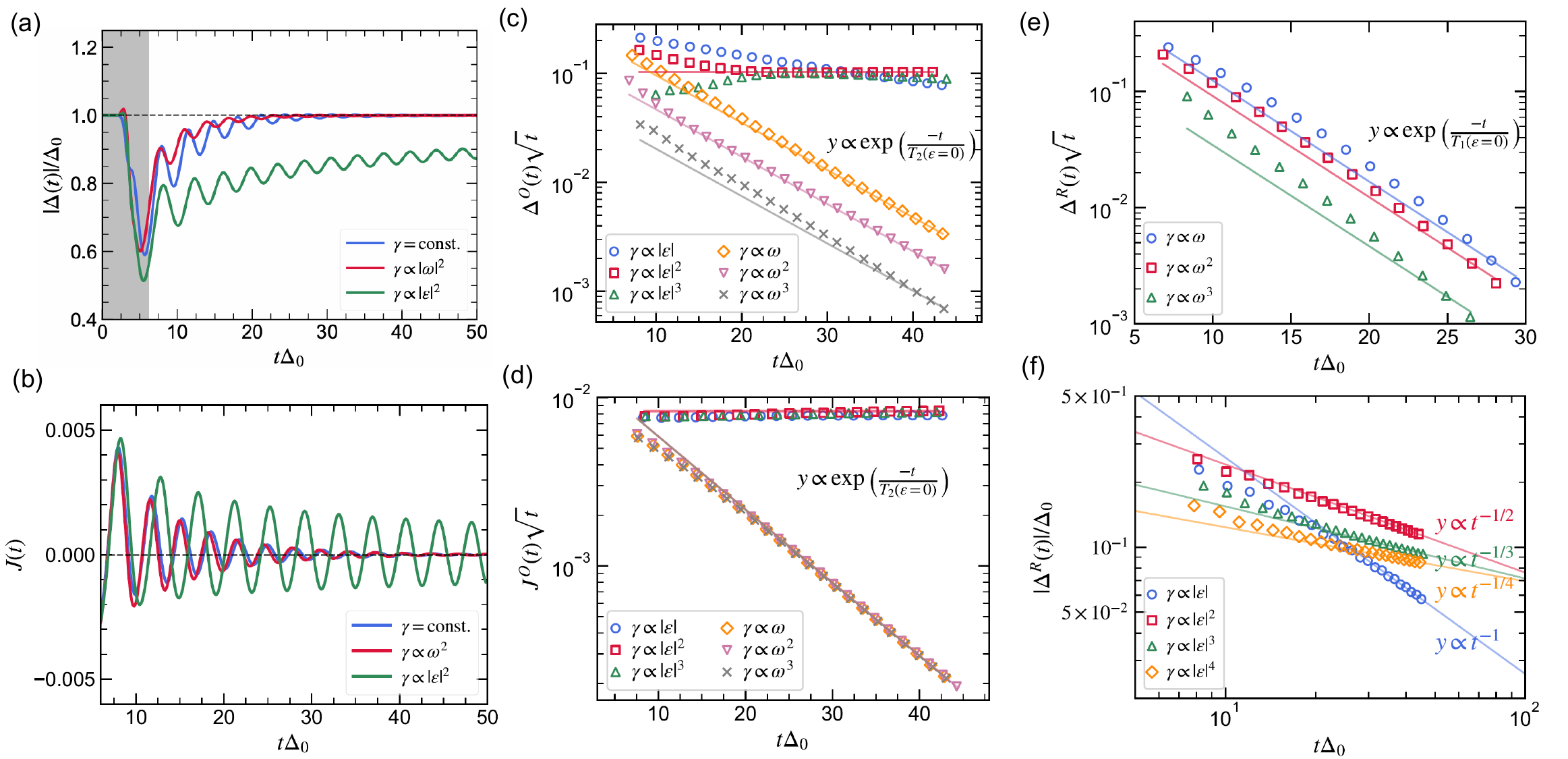}
    \caption{(a) Gap dynamics with different pseudospin damping forms under $x$-polarized pump pulse with intensity $\tilde A^2=4$ . The damping coefficient is chosen to be $\gamma_0=0.1$. (b) Nonlinear current $J(t)=J_{xxxx}(t)$ with different pseudospin damping forms.
    (c) Oscillations of the  Higgs mode $\Delta^O(t) $ and (d) the nonlinear current $J^O(t)$ multiplied by $\sqrt{t}$ in a half-log plot. For $\gamma\propto \omega^p$, the
    exponential decay allows us to extract pseudospin dephasing: $1/T_2(\varepsilon=0)$. (e) Gap recovery multiplied by $\sqrt{t}$ for $\gamma\propto \omega^p$, allowing extraction of quasiparticle redistribution: $1/T_1(\varepsilon=0)$. (f) Gap recovery for $\gamma\propto |\varepsilon|^p$ shows power-law $t^{-1/p}$.} 
    \label{fig:2}
\end{figure*}

Figure \ref{fig:2}(a) shows the gap dynamics under $x$-polarized light for different types of damping with $\gamma_0=0.1$. As expected, the presence of damping modifies the gap dynamics in two ways: (1) a gradual recovery of the gap $\Delta^R(t)$ towards its equilibrium value, and (2) a reduction in the oscillation amplitude $\Delta^O(t)$. For the relaxation of the nonlinear current, we mostly observe the oscillation decay $J^O(t)$ rather than the recovery rate $J^R(t)$ since $J(t)$ is oscillating around the equilibrium value, as shown in Fig.\;\ref{fig:2}(b).

\subsubsection{Oscillatory decay}

The amplitude oscillation originates from the transverse components of the pseudospins and can be expressed as

\begin{equation}
      O(t) \sim \int_{-\infty}^{\infty} d\varepsilon \, f^O(\varepsilon) \cos[\omega(\varepsilon) t] e^{-\gamma(\varepsilon)t}.
\end{equation}
where the pseudospin dephasing rate $\gamma(\varepsilon)= 1/T_2(\varepsilon) $.

\textbf{(1) Homogeneous damping:}  
The amplitude oscillation obeys $O(t)\propto \frac{\cos(2\Delta_{t} t + \pi/4)}{\sqrt{t}} e^{-\gamma_0 t}$.  

\textbf{(2) Inhomogeneous} $\bm{\gamma(\varepsilon) = \gamma_0 (\omega_k/2)^p}$:  
By applying the saddle-point method, the condition 
$\frac{d}{d\varepsilon}\left[i\omega(\varepsilon) - \gamma(\varepsilon)\right] = 0$ is satisfied at $\varepsilon = 0$. Expanding around this critical point and evaluating the integral using the Gaussian approximation, the long-time oscillation becomes

\begin{equation}
      O(t) \sim f^O(\varepsilon=0)\frac{\cos \left( 2\Delta_{t} t + \tfrac{1}{2}\tan^{-1}\!\left(\tfrac{\omega_2}{\gamma_2}\right)\right)}{\sqrt{t(\omega^2_2+\gamma_2^2)^{1/2}}} e^{-\gamma(\varepsilon=0)t},
\end{equation}
where $\omega_2 = \omega''(\varepsilon=0)$ and $\gamma_2=\gamma''(\varepsilon=0)$. Since $\omega_2 \gg \gamma_2$ in our setting, this reduces to

\begin{equation}
      O(t) \propto \frac{\cos(2\Delta_{t} t + \pi/4)}{\sqrt{t}} e^{-\gamma_0 t},
\end{equation}  
Observe that the signal is dominated by the decay rate at the Fermi surface. Figure\;\ref{fig:2}(c)(d) 
show the decay of Higgs amplitude $\Delta^O(t)$ and the current oscillation $J^O(t)$ multiplied by $\sqrt{t}$ in a half-log plot. With this correction, the macroscopic observable decays exponentially, and the pseudospin dephasing rate on the Fermi surface can thus be extracted from the long-time dynamics with $1/T_2(\varepsilon=0)=\gamma_0$. 

\textbf{(3) Inhomogeneous} $\bm{\gamma(\varepsilon) = \gamma_0 |\varepsilon|^p}$:  
For $p=1$, the saddle-points lie in the complex plane, $z_0 = \pm i\frac{\gamma_0\Delta_t}{\sqrt{4+\gamma_0^2}}$, giving
\begin{equation}
      {O}(t) \propto \frac{1}{\sqrt{t}} \cos\!\left(\frac{2\Delta_{t}}{\sqrt{1+\gamma_0^2/4}} t + \pi/4\right) e^{-\frac{\gamma_0^2\Delta_t}{\sqrt{4+\gamma_0^2}} t}.
\end{equation}
Thus, the oscillation acquires a small additional exponential damping, with an exponent approximately $\gamma_0^2/2$.
For $p\ge 2$, the saddle point lies on the Fermi surface $\varepsilon=0$, yielding
\begin{equation}
 {O}(t)  \propto \frac{1}{\sqrt{t}} \cos(2\Delta_{t} t + \pi/4),
\end{equation}
Therefore, if the pseudospin dephasing $1/T_2 = \gamma(\varepsilon)$ vanishes sufficiently rapidly approaching the Fermi surface, {macroscopic observables} would decay as a pure power-law, much as in the undamped case. 

\subsubsection{Recovery rates}

The recovery signal is mostly observed in the gap dynamics and it arises from the relaxation of the longitudinal components. For s-wave superconductors, we write

\begin{equation}
    \Delta^R(t) \sim \sum_k f^R_k |\delta s_k^\parallel(t)| \sim \int_{-\infty}^{\infty} d\varepsilon \, f^R(\varepsilon) e^{-2\gamma(\varepsilon)t}.
\end{equation}
where the quasiparticle redistribution rate is $2\gamma(\varepsilon)= 1/T_1(\varepsilon)$.

\textbf{(1) Homogeneous damping:}  
The recovery rate is $\Delta^R(t)\propto e^{-2\gamma_0 t}$.  

\textbf{(2) Inhomogeneous} $\bm{\gamma(\varepsilon) = \gamma_0 (\omega_k/2)^p}$:  
By Laplace’s method for $t\to\infty$,  

\begin{equation}
        \Delta^R(t) \propto \frac{e^{-2\gamma_0 t}}{\sqrt{t}}.
\end{equation}
In Fig.~\ref{fig:2}(e), we multiply the gap recovery by $\sqrt{t}$ and plot its decay on a half-log scale, showing that the quasiparticle relaxation rate on the Fermi surface can be measured.  In our setting, $1/T_1(\varepsilon=0) = 2\gamma_0 = 0.2$.

\textbf{(3) Inhomogeneous} $\bm{\gamma(\varepsilon) = \gamma_0 |\varepsilon|^p}$:  

\begin{equation}
    \Delta^R(t) \propto \int_{-\infty}^{\infty} e^{-\gamma_0|\varepsilon|^p t} d\varepsilon \propto \frac{1}{t^{1/p}}.
\end{equation}
In contrast to the previous case, the recovery rate becomes a power-law decay. Most importantly, it is not related to the damping strength $\gamma_0$, but depends only on the energy dependence $p$. Figure \ref{fig:2} (f) shows the numerical results for for $p=1,2,3,4$. This implies that by measuring the long-time recovery, one can extract the energy dependence $p$ of the pseudospin relaxation.  

To conclude, the macroscopic signal is dominated by the pseudospin damping on the Fermi surface. When the damping on the Fermi surface is nonzero, the macroscopic signal exhibits an exponential decay with an additional $1/\sqrt{t}$ power law. The $T_1$ and $T_2$ times can be extracted from the gap recovery and oscillation decay respectively. If the psuedospin damping vanishes on the Fermi surface, then the gap recovery reverts to a pure power law, and it may be possible to extract the energy dependent damping rate from the power law exponent. 

\section{$d$-wave superconductors}
\label{sec: d-wave}
We now consider $d$-wave superconductors on a 2D square lattice with a single-particle band dispersion given by 
$\varepsilon_k= -2J(\cos{k_xa}+\cos{k_ya})-4J'\cos{k_xa} \cos{k_ya} - 2J''(\cos{2k_xa}+\cos{2k_ya}) - \mu$ ,
where the second- and third-nearest-neighbor hoppings are $J'=0.15J$ and $J''=0.5J'$, respectively. Including remote hoppings allows us to induce different irreps of the superconducting condensate, as discussed in previous sections.  Unlike the $s$-wave case, the energy gap has momentum dependence $\Delta_k=\Delta_0\cos2\phi_k$. Numerically, we choose $J/\Delta_0 = 10$ and set the Fermi surface at $\mu=0$ (which again is away from Van Hove singularity, given the presence of further neighbor hoppings). The pump is modeled as a Gaussian vector potential, as before. 

\begin{figure*}[htb]
    \centering
    \includegraphics[width=0.85\textwidth]{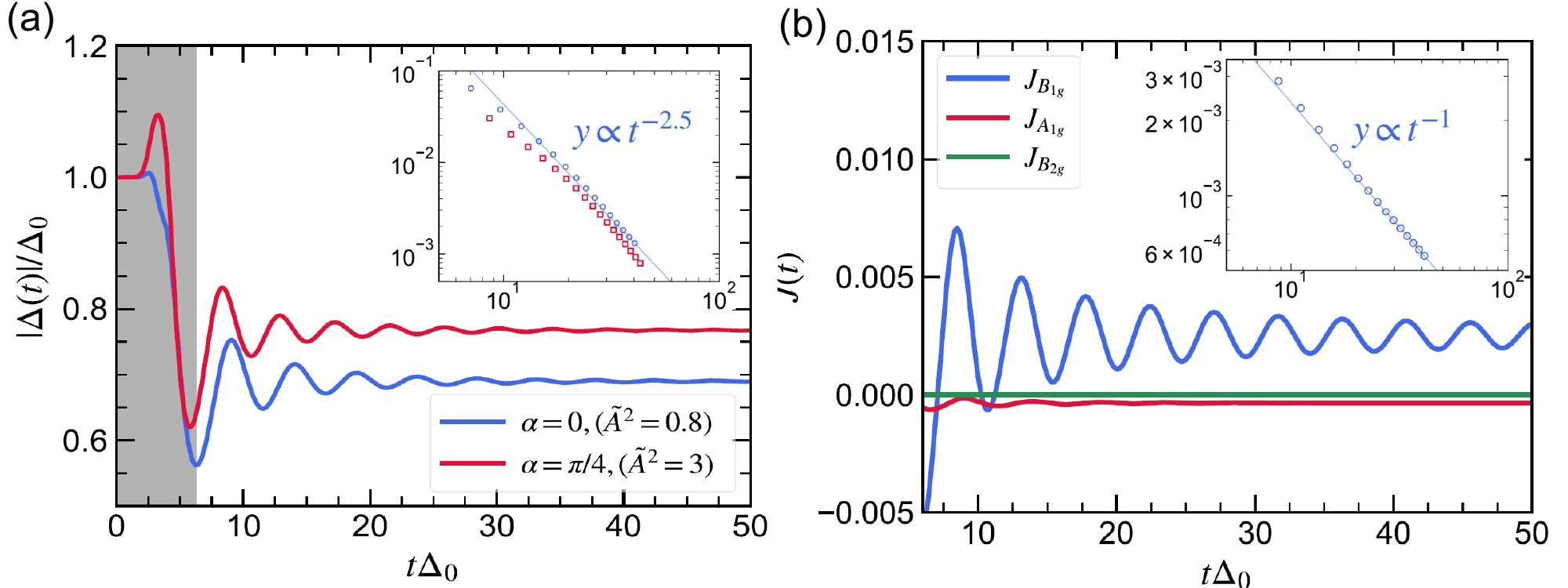}
    \caption{(a) Gap dynamics of d-wave superconductors under $x$ ($\alpha=0$) and $x'$ ($\alpha=\pi/4)$ polarized light with Fermi level $\mu=0$.  Inset: Higgs amplitude decay on a log-log scale, showing a much faster power-law decay than in s-wave superconductors. (b) Nonlinear current with different irreps induced by x-polarized pump pulse. The oscillation amplitude of $J_{B_{1g}}$ decays as $1/t$, as shown in the inset.  $J_{A_{1g}}$ has much weaker intensity and $J_{B_{2g}}=0$ because the x-polarized pump cannot excite the $B_{2g}$ irreps. }
    \label{fig:3}
\end{figure*}

\subsection{Collisionless limit}

For $d$-wave superconductors, only the $B_{1g}$ component of the pseudospins contributes to the gap, since $\Re[\delta\Delta(t)]=V_0\sum_k \cos(2\phi_k) \delta s^x_k $. However, using the nonlinear current, we can probe different irreps as discussed previously. 
The amplitudes of the gap and nonlinear current decay over time, just as in the $s$-wave case. At late times, the oscillatory part of these macroscopic observables can be expressed as:
\begin{align}
O(t)=\sum_k f_k e^{i\omega_k t}
    \sim\int_{-\infty}^{\infty} d\varepsilon  \int_{0}^{2\pi}d \phi  f(\varepsilon,\phi)e^{i\omega(\varepsilon,\phi)t}.
\end{align}
Here we rewrite the summation as an integration over $\varepsilon$ and $\phi$. There are four saddle points at the antinodal points $\phi=0,\pm \pi/2,\pi$ and four local minima at the nodal points $\phi=\pm \pi/4,\pm 3\pi/4 $ on the Fermi surface $\varepsilon=0$. Since the oscillation frequency at the nodal points vanishes, their contributions to the oscillation amplitude are negligible. At the antinodal points, the pseudospin precession frequency is $\omega=2\Delta$, corresponding to the Higgs mode energy. Expanding around the antinodal point $(\varepsilon,\phi)=(0,0) $:
\begin{equation}
    \omega = 2\Delta\left(1+\frac{\varepsilon^2}{2\Delta^2}-2{\phi^2}+\dots\right).
\end{equation}
If we further assume $f(\varepsilon,\phi)=f(\varepsilon)f(\phi)$, then the $\varepsilon$ and $\phi$ integrals can be treated separately using the stationary phase approximation to obtain:
\begin{align}
    O(t) &\sim  e^{i2\Delta t}\left[\int_{-\infty}^{\infty} d\varepsilon f(\varepsilon) e^{i\frac{\varepsilon^2 }{\Delta} t} \right]  \left[ \int_{0}^{2\pi}d \phi  f(\phi)e^{-i4\Delta \phi^2t} \right] \nonumber \\
&\propto f(\varepsilon=0)f(\phi=0)e^{i2\Delta t}/t.
\label{d-stationary}
\end{align}
In the collisionless limit, the oscillation is dominated by the antinodes and exhibits $1/t$ decay, faster than in the $s$-wave case due to the additional $\phi$-dependent dephasing of pseudospins. In other words, the intrinsic inhomogeneity of the pseudomagnetic field causes a faster decay of macroscopic observables in $d$-wave superconductors. 

Figure \ref{fig:3} shows numerical simulations of the gap dynamics and nonlinear current in the absence of pseudospin damping. After applying a pump pulse, the gap is quenched and undergoes coherent oscillations similar to those in s-wave superconductors. As illustrated in Sec.\;\ref{Pseudospin_texture}, the induced irreps are $\delta s^x_k=B_{1g}+A_{1g} $ for $x$-polairzed light and $\delta s^x_k =B_{1g}+A_{2g} $ for $x$'-polaized light. The $B_{1g}$ contribution dominates the gap dynamics since the other modes sum to zero in the gap equation. In Fig. \ref{fig:3}(b), the nonlinear current is induced by $x$-polarized light, and is mostly observed in the ${B_{1g}}$ channel, The nonlinear current corresponding to each irreducible representation can be obtained as follows:
\begin{align}
J_{B_{2g}}&= J_{xyxx} \nonumber \\ 
J_{A_{1g}} &=  J_{x'x'xx}-J_{B_{2g}} \\
J_{B_{1g}} &= J_{xxxx} - J_{A_{1g}} \nonumber
\end{align}
We observe that $B_{2g}$ mode cannot be excited by the x-polarized pump, as discussed in section\;\ref{irreps}. The $A_{1g}$ mode is much smaller than $B_{1g}$ , such that $J_{xxxx}\approx J_{B_{1g}}$, which shows coherent oscillation similar to the Higgs amplitude mode. 
The oscillation amplitude decay of the Higgs mode and  $J_{B_{1g}}$ is shown in the inset. In our numerical results, the Higgs amplitude mode decays approximately as  $t^{-2.5}$ , which is much faster than our prediction of $1/t$. This discrepancy may arise because the in-plane component of pseudospin dephasing has a more complicated distribution or dynamics such that the approximation in Eq.\; \ref{d-stationary} does not strictly hold.  A similarly fast decay was also observed for interaction quenches in \cite{PSC}. Nevertheless, we find that the amplitude decay of the nonlinear current $J_{B_{1g}}$ is well fitted by $1/t$, consistent with the stationary phase approximation. Our results point out that in $d$-wave superconductors, the Higgs mode and the nonlinear current exhibit different decay rates even in the collisionless case. The different decay rates of these two quantities originate from their distinct physical origins: in clean superconductors, the nonlinear current is dominated by quasiparticle contributions rather than by the Higgs mode.

 \begin{figure*}[htb]
    \centering
    \includegraphics[width=0.85\textwidth]{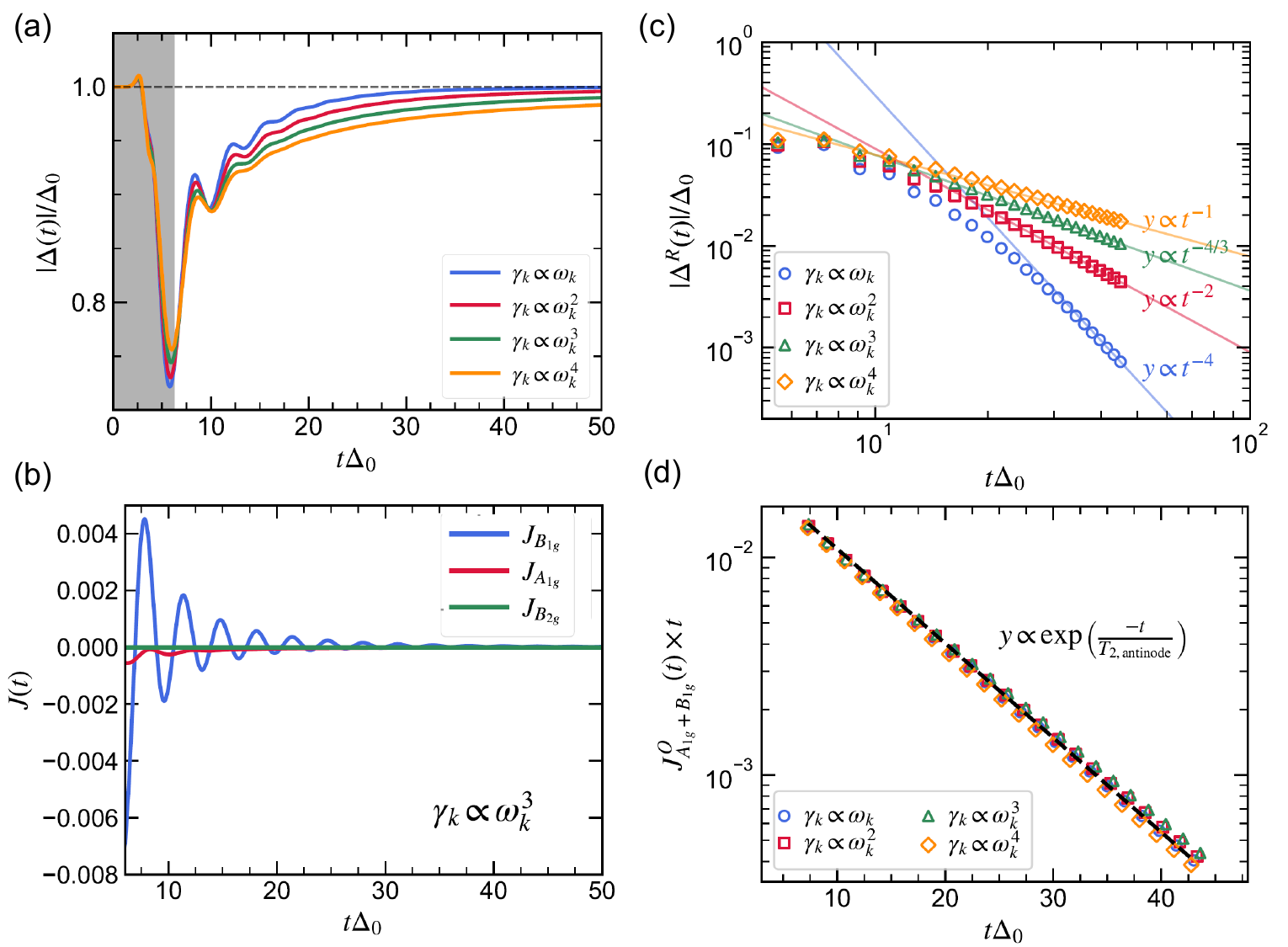}
    \caption{Gap dynamics  for d-wave superconductors with pseudospin damping forms $\gamma_k = \gamma_0(\omega_k/2)^p$with $\gamma_0=0.1$. In all figures, we use x-pump pulse with intensity $\tilde A^2=0.8$. (b) Nonlinear current with $\gamma_k\propto\omega^3_k$. (c) Gap recovery is dominated by the antinodes and decays as $t^{-4/p}$. (d) Decay of the oscillation amplitude of $J_{xxxx}=J_{A_{1g}+B_{1g}}\approx J_{B_{1g}}\propto e^{-t/T_{2,antinode}}/t$, which is valid for different pseudospin damping forms.}
    \label{fig:4}
\end{figure*}

\subsection{Pseudospin relaxation}
As before, we introduce quasiparticle relaxation time $T_1$ and pseudospin dephasing time $T_2$.  For $d$-wave superconductors, quasiparticles' relaxation can exhibit momentum dependence. We consider homogeneous damping type $\gamma_k=\gamma_0$ and the inhomogeneous (energy-momentum dependent) damping as $\gamma_k=\gamma_0 (\omega_k/2)^p=\gamma_0 [\varepsilon_k^2+{\Delta}^2_k]^{p/2}$. These relaxation types match at the antinodes with $\gamma_k=\gamma_0$. On the Fermi surface, the inhomogeneous relaxation rate varies within the range  $0\leq \gamma_k=\gamma_0\Delta_k=\gamma_0 \Delta_0|\cos(2\phi_k)| \leq \gamma_0 $, where we set $\Delta_0=1$ for convenience.

For the oscillatory part, we can write
\begin{equation}
     O(t) \sim    \int_{-\infty}^{\infty} d\varepsilon\int_{0}^{2\pi}d \phi  f(\varepsilon,\phi)e^{i\omega(\varepsilon,\phi)t}  e^{-\gamma(\varepsilon,\phi)t}.
\end{equation}
where $\gamma(\varepsilon,\phi)=1/T_2(\varepsilon,\phi)$

\textbf{(1) Homogeneous damping:} $O(t)\propto \cos({2\Delta_tt}) e^{-\gamma_0  t}/t^b $
where $b$ equals to the power-law exponent in the collisionless limit. 

\textbf{(2) Inhomogeneous} $\bm{\gamma(\varepsilon) = \gamma_0 (\omega_k/2)^p}$. The saddle point is found at the antinodes, and we thus obtain
$O(t)\propto \cos({2\Delta_tt}) e^{-\gamma_0t}/t^b $ at late times, consistent with the homogeneous case. Here, the decay rate can be written as $\gamma_0=\gamma_{antinode}= 1/T_{2,antinode}$.
We do not explicitly write the value of coefficient $b$ due to the fact that it depends on the measured quantity. For Higgs amplitude oscillation, we estimate $b\approx2.5$, while for the nonlinear current amplitude $J_{xxxx}\approx J_{B_{1g}}$,  we have shown $b=1$. 

Figure \ref{fig:4} shows the dynamics of gap and nonlinear current under pseudospin relaxation $\gamma_k\propto \omega^p_k$. We find that the gap oscillation amplitude at late times becomes invisible especially for large $p$. Thus, a practical way to extract $T_2$ would be measuring the oscillation decay of the nonlinear current $J_{xxxx}$, which decays as $1/t$ in the collisionless limit. The amplitude of $J$ after correcting the  $1/t$ decay is plotted  in Figure \ref{fig:4}(d) . It turns out that the nonlinear current $J_{xxxx}$ indeed decays as $e^{-t/T_{2,antinode}}/t$, from which we can extract $T_{2}$ at the antinodes.

On the other hand, the recovery rate can be written as
\begin{equation}
     R(t) \sim    \int_{-\infty}^{\infty} d\varepsilon\int_{0}^{2\pi}d \phi  f^R(\varepsilon,\phi) e^{-2\gamma(\varepsilon,\phi)t}
\end{equation}
where $2\gamma(\varepsilon,\phi)=1/T_1(\varepsilon,\phi)$ in our setting. 

\textbf{(1) Homogeneous damping:} For homogeneous damping, the recovery rate would decay as $e^{-2\gamma_0t}$. 

\textbf{(2) Inhomogeneous} $\bm{\gamma(\varepsilon) = \gamma_0 (\omega_k/2)^p}$: Using the Laplace's method, the recovery rate should be dominated by the nodal points. Since the  pseudospin damping at the nodal points vanishes, we expect a power-law decay. One might think that this results in a $t^{-2/p}$ decay ($t^{-1/p}$ from $\varepsilon$ integral and $t^{-1/p}$ from $\phi$ integral). But we need to be careful that the pseudospins actually vanish at the nodes (ill defined). %Mathematically, if consider 
%\begin{align*}
%R_{s,n,k,p}(t) &= \int_{0}^{2\pi} d\phi \int_{-\infty}^{\infty} d\varepsilon\;
%\frac{|\varepsilon|^{s} \, |\cos(2\phi)|^{n}}{|\omega(\varepsilon,\phi)|^{k}} \,
%e^{-\gamma_0\omega^{p}(\varepsilon,\phi) t} \\
%&\propto t^{\frac{s+n-k+2}{p}}
%\end{align*}
%It turns out that the power depends on how fast the prefactor vanishes approaching the nodes. It might be reasonable to choose $s=n=k=2$ (assumption, need to check \RN{remind me where these numbers come from?}),
Numerically, we find that the gap recovers as $\Delta^R(t)\propto t^{-4/p}$ as shown in Fig.\ref{fig:4}(c). We conclude that it is plausible to extract the energy-momentum dependence of pseudospin relaxation, characterized by the power $p$, from the gap recovery rate.

\begin{figure*}[htb]
    \centering
    \includegraphics[width=0.85\textwidth]{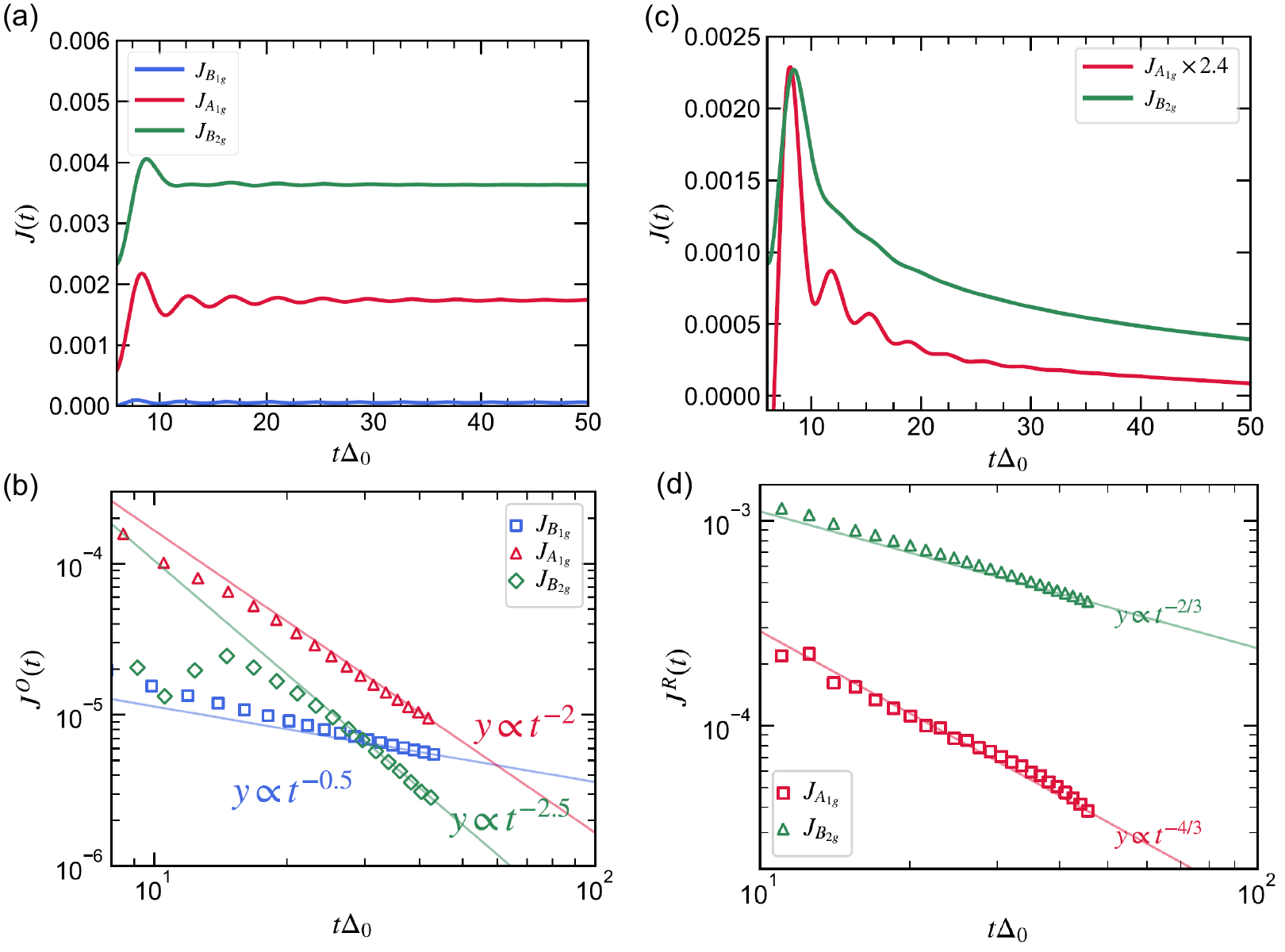}
    \caption{(a) Nonlinear current induced by an $x'$-polarized pump pulse with intensity $\tilde A^2=3$ in the collisionless limit, resolved into different irreducible representations. 
    (b) Oscillation amplitudes in the collisionless limit, showing that the $B_{2g}$ mode decays the fastest. 
    (c)--(d) Nonlinear current with pseudospin damping $\gamma_k \propto \omega_k^3$, where the $B_{2g}$ mode decays more slowly.}
    \label{fig:5}
\end{figure*}

\subsection{Light polarization effect}

 In  references \cite{Papenkort_coherent, schwarz_classification_2020}, the light coupling to $d$-wave superconductor is treated by considering a Hamiltonian 
$H_{\text{EM}}= \frac{e\hbar}{2m} \sum_{\mathbf{k},\mathbf{q},\sigma} (2\mathbf{k} + \mathbf{q}) \cdot \mathbf{A}_{\mathbf{q}}(t) \, c^{\dagger}_{\mathbf{k}+\mathbf{q},\sigma} c_{\mathbf{k},\sigma} 
\quad + \frac{e^2}{2m} \sum_{\mathbf{k},\mathbf{q},\sigma} \left( \sum_{\mathbf{q}'} \mathbf{A}_{\mathbf{q}-\mathbf{q}'}(t) \cdot \mathbf{A}_{\mathbf{q}'}(t) \right) c^{\dagger}_{\mathbf{k}+\mathbf{q},\sigma} c_{\mathbf{k},\sigma}
$
where $\boldsymbol{q}$ is the photon momentum. Reference \cite{schwarz_classification_2020} points out that it is possible to excite different modes of the condensate oscillation by changing the direction of the photon momentum $\boldsymbol{q}$. They thus conclude that $\boldsymbol{q}\ne0$ is an important condition to excite different irreps of supercondcuting condensate. 

In this work, instead, we focus on the $\boldsymbol{q}=0$ limit. In this case, the linear coupling (first term) vanishes, leaving only the quadratic term. We thus treat the light-matter interaction using the minimal coupling scheme at second order in the vector potential. Importantly, we argue that different irreps can indeed be excited even at $\boldsymbol{q}=0$ simply by changing the light polarization. Once excited, these irreps can be detected separately in the nonlinear current by choosing different combinations of probe polarization and current measurement directions.

In the previous section, we focused on an $x$-polarized pump, which predominantly excites the $B_{1g}$ mode. Here, we consider an $x'$-polarized pump ($\alpha=\pi/4$), where both $B_{2g}$ and $A_{1g}$ modes of the psuedospins' $z$-component can be significantly excited, as shown in Fig.~\ref{fig:5}(a). In the collisionless limit, the oscillation amplitudes of the nonlinear currents decay as power laws, as shown in Fig.~\ref{fig:5}(b). Among the different irreps, $J_{B_{2g}}$ decays the fastest. This can be explained by our previous argument: the oscillation is dominated by contributions from the antinodes, while the $B_{2g}$ mode vanishes at the antinodes, leading to a faster decay.

When pseudospin relaxation is included with $\gamma_k\propto\omega_k^3$, the nonlinear current dynamics are shown in Fig.~\ref{fig:5}(c). We normalize the magnitude of $J_{A_{1g}}$ in comparison with $J_{B_{2g}}$. In contrast to the oscillatary decay, the recovery of $J_{B_{2g}}$ is slower than that of the $J_{A_{1g}}$ mode because the recovery rate is dominated by the nodes instead. Their long-time recovery follows a power-law decay, as demonstrated in Fig.~\ref{fig:5}(d).

% \textit{Note:} What is not shown here is the recovery of the $J_{B_{1g}}$ mode, which is much smaller in magnitude than the other modes. However, our numerical results indicate that it decays as $t^{-2/3}$, the same as the $J_{B_{2g}}$ mode. At present, we do not have an analytic explanation for this numerical observation.

To summarize: we can selectively probe the relaxation of modes in different irreps by altering the pump and probe polarization, as well as altering the direction in which we measure nonlinear current. In the absence of intrinsic pseudospin damping, the oscillation decay of macroscopic observables are a power law function of time, with larger exponent than in the $s$-wave case. In the presence of intrinsic damping the oscillations decay exponentially at a rate controlled by the damping at the antinodes, whereas the recovery rates are controlled by the damping at the nodes, and follows either an exponential or power law time dependence depending on the nature of the damping. Thus, the intrinsic damping at the nodes and antinodes may therefore be extracted from the recovery and oscillation decay respectively.

% . , the oscillation amplitude decays but faster than $1/t$. The $J_{B_{2g}}$ decays fastest compared to other channels. When the pseudosin realxation is present   Fig.\ref{fig:5}(b)., the nonlinear current has a recovery rate and faster amplitude decays. For the recovery rate, we observe that the $J_{B_{2g}}$
%  decays faster in short times, but eventually becomes similar as $J_{A_{1g}+B_{2g}}$ at late times as expected because the $A_{1g}$ channel decays faster.

% The damping behaviour of recovery and oscillation at late times can be understood by the same argument before. The recovery rate (at late times) is determined by the nodes, and thus depends on whether the prefactor vanishes near the nodes. Since only $B_{1g}$  vanish at the nodes, we have the decay rate: $J_{B_{1g}}>J_{A_{1g}}\approx J_{B_{2g}}$. For the oscillation, the situation is reversed because it is determined by the antinodes, where only the $B_{2g}$ channel vanishes, thus the recovery rate  $J_{B_{2g}}>J_{A_{1g}}\approx J_{B_{1g}}$. This explain the result we obtain numerically. 

\section{Summary}
\label{sec: conclusions}
In this work, we demonstrate how the pseudospin relaxation times $T_1$ and $T_2$ can be extracted from macroscopic observables such as the superconducting gap and the nonlinear current. For $s$-wave superconductors, if the pseudospin damping rate on the Fermi surface is finite (e.g., $\gamma(\varepsilon)\propto \omega^p$), then the oscillations of the nonlinear current and the Higgs amplitude decays as $e^{-t/T_2(\varepsilon=0)}/\sqrt{t}$ and the gap recovery decays as $e^{-t/T_1(\varepsilon=0)}/\sqrt{t}$. Therefore, one can measure $T_1$ and $T_2$ by correcting for the intrinsic $1/\sqrt{t}$ decay that already appears in the collisionless limit due to inhomogeneity of the pseudomagnetic field. If, instead, the pseudospin relaxation vanishes on the Fermi surface (as at $T=0$), for example $\gamma(\varepsilon)\propto |\varepsilon|^p$ (with $p>1$), then the oscillations of macroscopic observables decay as $1/ \sqrt{t}$, and the gap recovery follows a power-law decay $t^{-1/p}$, allowing direct extraction of the energy-dependent relaxation exponent $p$. Numerically, we adopt a spin-length–preserving model with $1/T_2 = 1/(2T_1)$ at late times where no pure dephasing is introduced ($1/T_2^* = 0$). Nevertheless, the framework is general and can be extended to cases with pure dephasing.

For $d$-wave superconductors, the situation is more subtle. In the collisionless limit, the Higgs amplitude decays as $1/t^b$ with $b\approx 2.5$ (numerically), whereas the nonlinear current $J_{xxxx}$ decays as $1/t$. When pseudospin relaxation is introduced, the oscillation amplitude and the recovery rate become dominated by contributions from the antinodes and nodes, respectively. Assuming a relaxation of the form $\gamma_k \propto \omega_k^p$, which vanishes at the nodes but not at the antinodes, the oscillation amplitude acquires an exponential decay, while the recovery rate remains a power law. In practice, $T_{2,\text{antinode}}$ can be extracted from the nonlinear current $J_{xxxx}\propto e^{-t/T_{2,\text{antinode}}}/t$, while the gap recovery follows $\Delta^R(t)\propto t^{-4/p}$, providing access to the energy–momentum dependence of pseudospin relaxation.
We also study the role of light polarization. In particular, we find that the $B_{2g}$ pseudospin component can be excited by an $x'$-polarized pump (at $45^\circ$) and detected through the nonlinear current $J_{xyx'x'}$. By comparing decay rates for different irreps, both in the collisionless limit and with pseudospin damping, we confirm that distinct irreducible representations indeed exhibit different decay behaviors, and that these can be probed by adjusting pump and probe polarization, as well as the direction of nonlinear current probed. We hope that the results contained herein will prove useful for optical interrogation of clean superconductors. 

Finally, we note that disorder may affect the dynamics of superconductors. For instance, it has been discussed that in $s$-wave superconductors the Higgs mode may become undamped 
in the collisionless limit within a finite time window in the presence of magnetic impurities \cite{Li_2024, Dzero_2024}. For $d$-wave superconductors, disorder may wash out the oscillation peak of the nonlinear current at frequency $\omega=2\Delta$ \cite{Benfatto_2023}. How to disentangle disorder-induced decay rate from intrinsic relaxation rates with energy or momentum dependence, and what role polarization control plays in disordered superconductors,  remain open questions and are left for future work.

% How disorder affects the dynamics of intrinsic relaxation rates with energy or momentum dependence and polarization control remains an open question and is left for future work.}

{\bf Acknowledgements} We acknowledge useful discussions with Sarang Gopalakrishnan. We also thank Sarang Gopalakrishnan and Yang-Zhi Chou for feedback on the manuscript. This work was supported by the U.S. National Science Foundation under NSF grant number DMR-2516302.

\section{Computation details}
We numerically solve Eqs.   \ref{differential} and \ref{self-consistent} self-consistently using a fourth-order Runge--Kutta method to minimize computational errors. The length of each pseudospin is conserved at every time step. The pseudospins are sampled in polar coordinates with $N_r \approx 200 J/\Delta_0$ and $N_\phi = 200$. For $J/\Delta_0 = 10^3$, this corresponds to a dense radial grid of $N_r = 2\times 10^5$. 
To further reduce the computation domain, we restrict the pseudospins to those near the Fermi surface with a cutoff $|\varepsilon - \mu| < 6\Delta_0$, corresponding approximately to the Debye frequency. With this cutoff and $J/\Delta_0 \ll 1$, the pseudospins are sampled on a ring rather than over the full Brillouin zone. For s-wave superconductors, we choose $\mu=-3J$, so that the Fermi surface is not distorted and remains circular in two dimensions. For d-wave superconductors, we consider the full band dispersion at $\mu=0$, for which the Fermi surface is still well approximated by a circle. %\RN{We should upload the computational scripts to Zenodo along with the submission.}

\section*{Data Availability}
All numerical codes used in this study are available at \href{https://doi.org/10.5281/zenodo.17393630}{10.5281/zenodo.17393630}.

\bibliography{Ref}

\end{document}